\documentclass[journal]{IEEEtran}

\usepackage{cite}

\ifCLASSINFOpdf
\usepackage{float} 
\usepackage{graphicx}
\usepackage{color}

\else
\fi
\usepackage{amsmath}
\usepackage[utf8]{inputenc}
\usepackage{multirow}
\usepackage{hyperref}
\usepackage[tableposition=top]{caption}
\usepackage{setspace}
\usepackage[linesnumbered,lined,commentsnumbered]{algorithm2e}

\begin{document}
%
\title{A 3D Motion Vector Database for Dynamic Point Clouds}
%
%
%

\author{André L. Souto,
        Ricardo L. de Queiroz,
        and Camilo Dorea 
\thanks{Use of the  database is free with the mandatory reference to the present document. We make no guarantees whatsoever about any of its contents.}
\thanks{A. L. Souto is with the Electrical Engineering Department at University of Brasilia, Brasilia, DF, Brazil, e-mail: andre@image.unb.br.}
\thanks{R. L. de Queiroz and C. Dorea are with the Computer Science Department at University
of Brasilia, Brasilia, DF, Brazil, e-mails: queiroz@ieee.org and camilodorea@unb.br.}
}

\maketitle

\begin{abstract}
Due to the large amount of data that point clouds represent \cite{overviewGraziozi:20} and the differences in geometry of successive frames, the generation of motion vectors for an entire point cloud dataset may require a significant amount of time and computational resources. With that in mind, we provide a 3D motion vector database for all frames of two popular dynamic point cloud datasets \cite{LoopC:16, dEonHMC:17}. The motion vectors were obtained through translational motion estimation procedure that partitions the point clouds into blocks of dimensions $M \times M \times M$, and for each block, a motion vector is estimated. Our database contains motion vectors for $M = 8$ and $M = 16$. The goal of this work is to describe this publicly available 3D motion vector database that can be used for different purposes, such as compression of dynamic point clouds. 

\end{abstract}

\begin{IEEEkeywords}
Database, Dynamic Point Clouds, 3D Motion Vector.
\end{IEEEkeywords}

\section{Introduction}

\IEEEPARstart{P}{oint} clouds consist of 3D images represented by a set of points in the 3D space. Each point has geometry, or coordinates ($x,y,z$), and attributes (e.g., colors) \cite{overviewGraziozi:20}. The point clouds used here are voxelized, that is, their points are constrained into small cubes with integer coordinates called voxels. A voxel is said to be occupied when it bounds one or more points. Successive frames in a point cloud sequence may present a different amount and location of occupied voxels, due to object motion. Because of this non-uniformity in geometry between point cloud frames, an adequate motion estimation in the 3D space is not trivial. Also, significant amount of time and computational resources may be demanded to estimate motion of several frames due to the large amount of data to represent point clouds \cite{DoreaQ:18, DoreaHQ:19}. 

With the aforementioned in mind, we provide a 3D motion vector database for dynamic point clouds. Our database is composed of 3D motion vectors for all frames of the popular Microsoft \cite{LoopC:16} and 8i \cite{dEonHMC:17} dynamic point clouds. They were obtained through block-based backwards motion estimation procedure of 3D scenes, therefore, its replication would demand a significant amount of time and computational resources.       

\section{Motion Estimation Procedure Used}
\label{sec:me}

In order to construct our database, we carried a block-based backwards motion estimation procedure to all frames in our test set. Let us consider the current frame to be estimated $\mathcal{I}(t)$ and the immediately preceding frame $\mathcal{I}(t-1)$. Initially, $\mathcal{I}(t)$ is partitioned into several non-overlapping blocks $\{b_{i}(t)\}$ of size $M \times M \times M$. Each block $b_{i}(t)$ is associated with a location in the space $(x_{b_{i} t}, y_{b_{i} t}, z_{b_{i} t})$ and is composed by a set of occupied voxels $\{v_{k}(t)\}$, where $k \leq M^{3}$. Each occupied voxel $v_{k}(t)$ is associated to geometry coordinates and color attributes $v_{k}(t) = [x_{v_{k} t}, y_{v_{k} t}, z_{v_{k} t}, Y_{v_{k} t}, U_{v_{k} t}, V_{v_{k} t}]$. In our database, we provide motion vectors for $M = 8$ and $M = 16$.

For each block $b_{i}(t)$, a block with the best match $b^{*}(t-1)$ is searched for in $\mathcal{I}(t-1)$. The search is performed within a search space defined by a block of dimensions $S \times S \times S$ centered at $(x_{b_{i} t}, y_{b_{i} t}, z_{b_{i} t})$. The search space may be represented by a list $\{b_{j}(t-1)\}$ of candidate blocks. For each block $b_{j}(t-1)$, voxel correspondences are determined among $\{v_{m}(t-1)\}$ and $\{v_{k}(t)\}$. For each $v_{k}(t)$, its nearest-neighbor in $\{v_{m}(t-1)\}$ is assumed to be its correspondence. After that, for each established correspondence, its Euclidean distance and color distance are computed. Then, the average Euclidean distance $\delta_{d}$ and the average color distance $\delta_{c}$ are determined. Finally, the block in $\{b_{j}(t-1)\}$ that presents the lowest cost $\delta$ (where $\delta = \delta_{d} + 0.35\delta_{c}$), is assigned as $b^{*}(t-1)$. Note that all possible displacements within the search space $S \times S \times S$ are tested in order to determine $b^{*}(t-1)$. The motion vector with components $(M_x, M_y, M_z)$ indicates the block in $\mathcal{I}(t-1)$ to be used to predict $b_{i}(t)$ as 

\begin{equation*}
\label{eq:mvs}
\begin{aligned}
&M_x = x_{b^{*} t-1} - x_{b_{i} t}, \\
& M_y = y_{b^{*} t-1} - y_{b_{i} t}, \\
& M_z = z_{b^{*} t-1} - z_{b_{i} t}.
\end{aligned}
\end{equation*}




In order to compensate the block $b_{i}(t)$ with the corresponding motion vector ($M_x, M_y, M_z$), the coordinates of $b^{*}(t-1)$ may be obtained by simply performing $(x_{b^{*} t-1}, y_{b^{*} t-1}, z_{b^{*} t-1}) = (x_{b_{i} t} + M_x, y_{b_{i} t} + M_y, z_{b_{i} t} + M_z)$. Then, all voxels inside $b^{*}(t-1)$ may have their coordinates subtracted by the motion vector $(x_{v_{m} t-1} - M_x, y_{v_{m} t-1} - M_y, z_{v_{m} t-1} - M_z)$.

\begin{table*}[!hbt]
\caption{Description of database contents.}
\label{tab:bd-gains}
\hspace*{-0.7cm}
\begin{tabular}{|lclcl|}
\cline{1-5}
\textit{\textbf{Point Cloud}} & \textit{\textbf{Depth}} & \textit{\textbf{Frames}} & \textit{\textbf{Block Size}} & \textit{\textbf{Filenames .mat}} \\ \hline
Andrew & 9         & 1-317  & 8  &  mvs\_andrew\_frame1\_d8.mat, \dots, mvs\_andrew\_frame317\_d8.mat           \\
Andrew & 9         & 1-317  & 16 &  mvs\_andrew\_frame1\_d16.mat, \dots, mvs\_andrew\_frame317\_d16.mat           \\
David & 9     & 1-215 & 8  & mvs\_david\_frame1\_d8.mat, \dots, mvs\_david\_frame215\_d8.mat   \\
David & 9     & 1-215 & 16  & mvs\_david\_frame1\_d16.mat, \dots, mvs\_david\_frame215\_d16.mat           \\
Phil & 9      & 1-244 & 8  & mvs\_phil\_frame1\_d8.mat, \dots, mvs\_phil\_frame244\_d8.mat   \\
Phil & 9     & 1-244 & 16   & mvs\_phil\_frame1\_d16.mat, \dots, mvs\_phil\_frame244\_d16.mat      \\
Ricardo & 9     & 1-215 & 8  & mvs\_ricardo\_frame1\_d8.mat, \dots, mvs\_ricardo\_frame215\_d8.mat       \\
Ricardo & 9     & 1-215 & 16  & mvs\_ricardo\_frame1\_d16.mat, \dots, mvs\_ricardo\_frame215\_d16.mat       \\
Sarah & 9    & 1-206 & 8   & mvs\_sarah\_frame1\_d8.mat, \dots, mvs\_sarah\_frame206\_d8.mat     \\
Sarah & 9    & 1-206 & 16   & mvs\_sarah\_frame1\_d16.mat, \dots, mvs\_sarah\_frame206\_d16.mat       \\
Longdress & 10  & 1052-1350 & 8   & mvs\_longdress\_frame1052\_d8.mat, \dots, mvs\_longdress\_frame1350\_d8.mat       \\
Longdress & 10  & 1052-1350  & 16  & mvs\_longdress\_frame1052\_d16.mat, \dots, mvs\_longdress\_frame1350\_d16.mat        \\
Loot & 10    & 1001-1299 & 8  & mvs\_loot\_frame1001\_d8.mat, \dots, mvs\_loot\_frame1299\_d8.mat \\
Loot & 10    & 1001-1299 & 16  & mvs\_loot\_frame1001\_d16.mat, \dots, mvs\_loot\_frame1299\_d16.mat    \\
Redandblack & 10   & 1451-1749 & 8  & mvs\_redandblack\_frame1451\_d8.mat, \dots, mvs\_redandblack\_frame1749\_d8.mat        \\
Redandblack & 10   & 1451-1749 & 16  & mvs\_redandblack\_frame1451\_d16.mat, \dots, mvs\_redandblack\_frame1749\_d16.mat          \\
Soldier & 10     & 537-835 & 8  & mvs\_soldier\_frame0537\_d8.mat, \dots, mvs\_soldier\_frame0835\_d8.mat    \\
Soldier & 10     & 537-835 & 16  & mvs\_soldier\_frame0537\_d16.mat, \dots, mvs\_soldier\_frame0835\_d16.mat       \\
 \hline
\end{tabular}
\end{table*}

\section{3D Motion Vector Database}
\label{sec:mv}
Our database provides 3D motion vectors for all frames of the point cloud sequences ``Andrew'', ``David'', ``Phil'', ``Ricardo'', ``Sarah'', ``Longdress'', ``Loot'', ``Redandblack'' and ``Soldier'' \cite{LoopC:16, dEonHMC:17} with depths 9 and 10. Figure 1 shows projections of a random frame of the point cloud sequences used. Table I presents the contents of the database. It is available at \url{https://queiroz.divp.org/data/motionVectorDatabase.zip}, website hosted by the Digital Image and Video Processing (DIVP) group at University of Brasilia. The motion files are provided in .mat format to be used in Octave or Matlab but can also be saved into .txt files and used with any other programming languages. They are named as in this following example: frame \#1065 of point cloud ``Longdress'' and blocks of dimension $8 \times 8 \times 8$, is named \textit{mvs\_longdress\_frame1065\_d8.mat}. Each file contains three variables: 

\begin{itemize}
 \item \textit{cubeDim}: represents the size of the blocks (8 or 16);
 \item \textit{cubeList}: contains the coordinate values (divided by the value of \textit{cubeDim}) of all cubes with at least one occupied voxel.
 \item \textit{MVList}: corresponds to the list of motion vectors ($x,y,z$), in the same order as variable \textit{cubeList};
\end{itemize}

\section{Conclusion}
\label{sec:cc}
We provide a 3D motion vector database for popular dynamic point clouds. The motion vectors are obtained through block-based backwards translational motion estimation. The goal of this work is to describe a publicly available 3D motion vector database, whose replication would demand significant amount of time and computational resources. It can be used for different purposes, one example being the compression of dynamic point clouds.

\begin{figure}[h]
\centering
\hspace*{0.10cm}
\includegraphics[width=9cm]{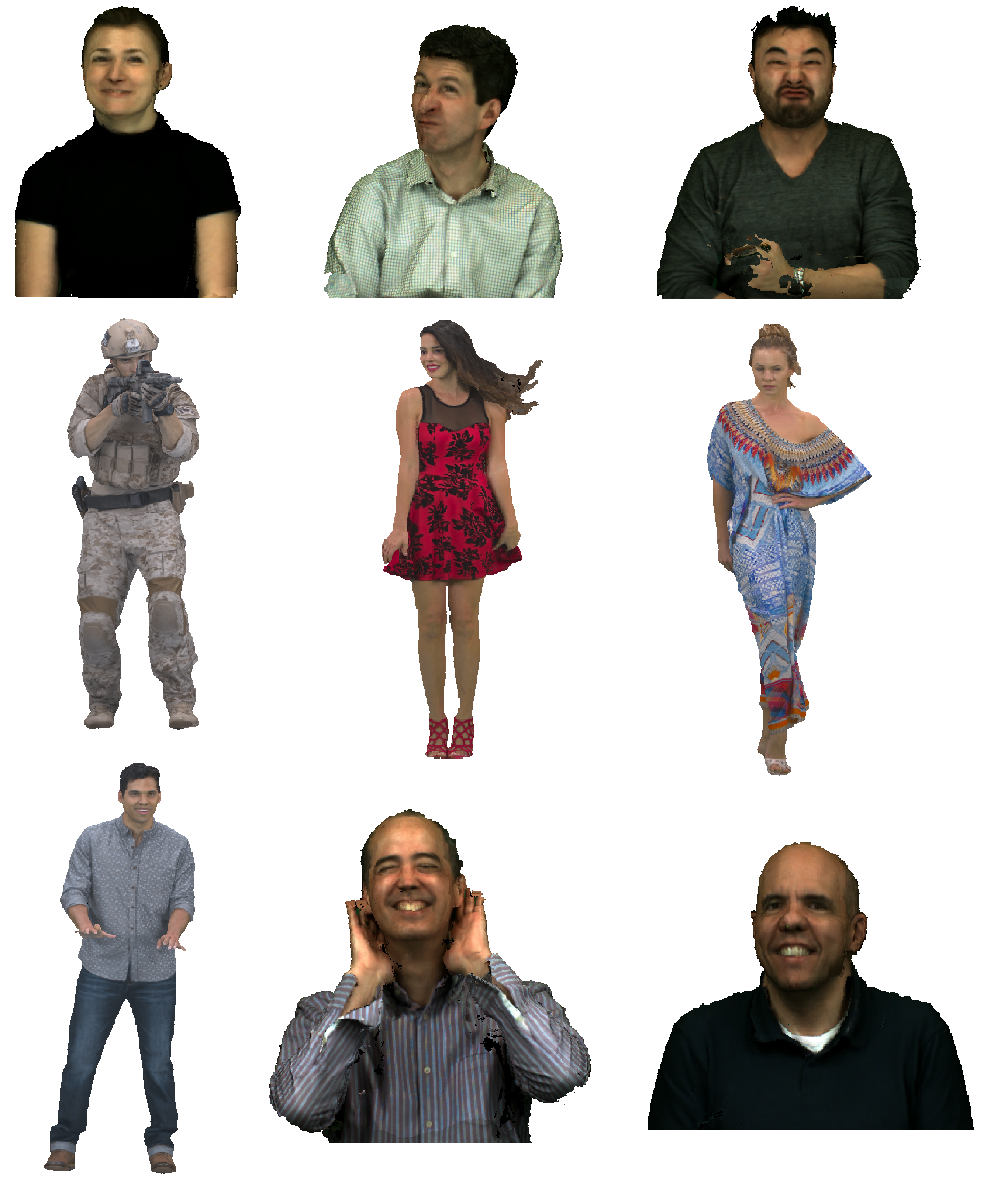}
\caption{Projections of point cloud sequences used. From left to right, top to bottom: ``Sarah'', ``Andrew'', ``David'', ``Soldier'', ``Redandblack'', ``Longdress'', ``Loot'', ``Phil'' and ``Ricardo''.}
\end{figure}

%
\IEEEpeerreviewmaketitle

\ifCLASSOPTIONcaptionsoff
  \newpage
\fi

\bibliographystyle{IEEEtran}
\bibliography{IEEEabrv,IEEEexample}

\end{document}